# Combined Raman scattering and *ab initio* investigation of pressure-induced structural phase transitions in the scintillator ZnWO$_4$

D. Errandonea[1,*], F.J. Manjón[2], N. Garro[3], P. Rodríguez-Hernández[4], S. Radescu[4], A. Mujica[4], A. Muñoz[4], and C.Y. Tu[5]

[1]MALTA Consolider Team, Departamento de Física Aplicada - ICMUV, Universitat de València, Edificio de Investigación, c/Dr. Moliner 50, 46100 Burjassot, Valencia, Spain

[2]MALTA Consolider Team, Departamento de Física Aplicada - IDF, Universitat Politècnica de València, 46022 València, Spain

[3]Fundació General de la Universitat de València-ICMUV, Universitat de València, Polígon La Coma s/n, 46980 Paterna (Spain)

[4]MALTA Consolider Team, Departamento de Física Fundamental II, Universidad de La Laguna, La Laguna 38205, Tenerife, Spain

[5]Fujian Institute of Research on the Structure of Matter, Chinese Academy of Sciences, Fuzhou, Fujian 350002, and Graduated School of Chinese Academy of Science, 100039 Beijing, China

**Abstract:** Room-temperature Raman scattering was measured in ZnWO$_4$ up to 45 GPa. We report the pressure dependence of all the Raman-active phonons of the low-pressure wolframite phase. As pressure increases new Raman peaks appear at 30.6 GPa due to the onset of a reversible structural phase transition to a distorted monoclinic β-fergusonite-type phase. The low- and high-pressure phases coexist from 30.6 GPa to 36.5 GPa. In addition to the Raman measurements we also report *ab initio* total-energy and lattice-dynamics calculations for the two phases. These calculations helped us to determine the crystalline structure of the high-pressure phase and to assign the observed

---

[*] Corresponding author, Email: daniel.errandonea@uv.es, Fax: (34) 96 3543146, Tel.: (34) 96 354 4475

Raman modes in both the wolframite and β-fergusonite phases. Based upon the *ab initio* calculations we propose the occurrence of a second phase transition at 57.6 GPa from the β-fergusonite phase to an orthorhombic *Cmca* phase. The pressure evolution of the lattice parameters and the atomic positions of wolframite ZnWO$_4$ are also theoretically calculated and an equation of state reported.

PACS NUMBERS: 62.50.+p, 63.20.-e, 78.30.-j

**I. Introduction**

Materials belonging to the tungstate family (AWO$_4$) have a long history of practical application, having been first used by Thomas A. Edison in 1896 to detect x-rays [1]. As a consequence of their technological importance, AWO$_4$ compounds have been the object of extensive research. The interest in them arises from their optical properties which form the basis of their wide application as phosphors, laser crystals, and scintillation detectors [2–4]. Recently, new applications for these materials have emerged, including large-volume scintillators for high-energy physics [5] and detectors devoted to the search of rare events (e.g. interactions with weakly-interactive massive particles) [6]. In particular, zinc tungstate (ZnWO$_4$), also known by its mineral name sanmartinite, is a wide-gap semiconductor, with band-gap energy close to 4 eV [7], and is a promising material for the new generation of radiation detectors [8].

It is well known that AWO$_4$ compounds mostly crystallize either in the tetragonal scheelite (space group (SG): *I4$_1$/a, Z=4*) or in the monoclinic wolframite (SG: P*2/c, Z=2*) structures depending on the size of the counter-cation A [9]. In particular, ZnWO$_4$ has a wolframite-type crystalline structure (see **Figure 1(a)**) [10], with two formula units (*Z*) per crystallographic cell. In this structure, both Zn and W cations have octahedral oxygen coordination and each octahedron shares two corners with its neighbours. In particular, the $WO_6$ octahedra are highly distorted since two of the W-O distances are much larger than the other four distances.

High-pressure research has proved to be an efficient tool to improve the understanding of the main physical properties of $AWO_4$ compounds. Although there is abundant literature on high-pressure studies in these materials, much of the research has been carried out on scheelite-structured compounds like $CaWO_4$, $SrWO_4$, $BaWO_4$ and $PbWO_4$ **[11 – 24]**. These studies have been recently reviewed **[25]** and have established that all the members of the scheelite subfamily of tungstates undergo a sequence of pressure-driven structural phase transitions with space-group changes $I4_1/a \rightarrow I2/a \rightarrow P2_1/n$ in good agreement with the conclusions drawn from the application of the Bastide's diagram **[26]** to $ABO_4$ compounds **[23, 25]**. Additionally, optical-absorption measurements on $PbWO_4$ **[27]** and luminescence studies on $SrWO_4$ **[28, 29]** showed that the electronic structure of scheelite tungstates is also strongly affected by pressure.

Given the structural differences between wolframite and scheelite **[9]**, their structures are expected to be modified in a different way upon compression **[30]**. However, very little information currently exists on how the crystal structures of $ZnWO_4$ and isostructural tungstates (e.g. $CdWO_4$ and $NiWO_4$) are affected by pressure. Only a couple of works reporting Raman measurements under pressure in $CdWO_4$ up to 40 GPa **[31]** and in $ZnWO_4$ up to 24 GPa **[32]** have been performed. Some contradictions and many unanswered questions arise from the information reported in these two works. For example, $ZnWO_4$ was found to remain stable in the wolframite structure up to 24 GPa **[32]**, while two phase transitions were reported in $CdWO_4$ at 10 GPa and 20 GPa **[31]**. In order to improve the knowledge of the physical properties of wolframite-type tungstates, as part of our project to study the structural stability of orthotungstates, we have carried out Raman spectroscopy measurements on ZnWO$_4$ up to 45 GPa. The obtained results are interpreted on the basis of first-principles total-energy and lattice-dynamics calculations. The technical aspects of the experiments and calculations are described in Sections II and III. The results are presented and discussed in Sec. IV. Finally, we present the conclusions of this work in Sec. V.

**II. Experimental details**

The samples used in the present experiments were obtained from a wolframite-type ZnWO$_4$ single crystal grown by the Czochralski method **[33]**. In order to get a high-quality crystal of ZnWO$_4$, the used raw materials such as ZnO and WO$_3$ must be of high purity. The raw materials used were ZnO (99.9%) and WO$_3$ (99.9%). The initial compounds were mixed in a carnelian bowl and sintered for almost 3 days at 1320 K. Then the charge was deposited in a Pt crucible bowl of ∅55x30 mm$^2$ and placed in a DJL-400 furnace. With the Pt wire rotating at a rate of 12 rpm and a pulling rate of 1.2 mm/h, the crystal was grown. When the procedure was over, the crystal was drawn out and cooled down to room temperature (RT) at a rate of 10–30 °C/h. The obtained crystal was optically transparent and colour free and x-ray diffraction measurements at ambient conditions showed that its diffraction pattern was in agreement with that of sanmartinite. The refined unit-cell parameters ($a$ = 4.680 Å, $b$ = 5.712 Å, $c$ = 4.933 Å, $\beta$ = 90.3°) and atomic positions compare well with earlier reported data from neutron powder-diffraction **[34]** (see **Table I**).

Two different samples were used for our Raman measurements under pressure. Sample # 1 was a 10 μm-thick plate cleaved along the {010} plane **[35]** from a ZnWO$_4$ single crystal. It was pressurized up to 21 GPa with a 16:3:1 methanol-ethanol-water mixture as pressure-transmitting medium. Sample #2 consisted in a pre-pressed pellet prepared using a finely ground powder obtained from the single crystal of ZnWO$_4$. It was pressurized up to 45 GPa with argon (Ar) to ensure better quasi-hydrostatic conditions **[36, 37]**. The pressure was determined by the ruby fluorescence technique **[38]** using the pressure scale recalibrated by Dewaele *et al.* **[39]**. RT Raman experiments were performed in backscattering geometry using the 488 nm (2.54 eV) line of an Ar$^+$-ion laser with a power of less than 100 mW before the DAC to avoid sample heating. Laser heating of the sample is negligible in the whole pressure range covered by our experiments because the laser energy is always below the band-gap of $ZnWO_4$ in both the low and high pressure phases. Note that in wolframite $ZnWO_4$ the band-gap is $E_g$ = 4 eV **[7]** and its pressure coefficient is positive ($dE_g/dP$ = 10 meV/GPa) **[40]**, and that $E_g$ is not expected to close more than 1 eV at the pressure-induced phase transition **[27]**. A Mitutoyo 20x long-working distance objective was employed for focusing the laser on the sample and for collecting the Raman spectra. The dispersed light was analyzed with a Jobin-Yvon T64000 triple spectrometer equipped with a confocal microscope in combination with a liquid nitrogen-cooled multi-channel CCD detector. The spectral resolution was better than 1 cm$^{-1}$ and Ar and He plasma lines were used to calibrate the Raman and photoluminescence spectra.

**III. Calculation technique**

Total-energy calculations and lattice-dynamics calculations were done within the framework of the density-functional theory (DFT) and the pseudopotential method using the Vienna *ab initio* simulation package (VASP) of which a detailed account can be found in **Ref. 41** and references therein. The exchange and correlation energy was initially taken in the generalized-gradient approximation (GGA) according to Perdew-Burke-Ernzerhof **[42]** (PBE) prescription. The projector-augmented wave (PAW) scheme **[43]** was adopted and the semicore 5*p* electrons of W were dealt with explicitly in the calculations. The set of plane waves used extended up to a kinetic energy cutoff of 500 eV. This large cutoff was required to deal with the O atoms within the PAW scheme to ensure highly converged results. The Monkhorst-Pack grid used for Brillouin-zone integrations ensured highly converged results (to about 1 meV per formula unit). We use 24 k-points, 26 k-points, 22 k-points and 6 k-points to study the wolframite, $CuWO_4$-type, β-fergusonite, and *Cmca* structures, respectively. At each selected volume, the structures were fully relaxed to their equilibrium configuration through the calculation of the forces on atoms and the stress tensor – see **Ref. 19**. In the relaxed equilibrium configuration, the forces are less than 0.002 eV/A and the deviation of the stress tensor from a diagonal hydrostatic form is less than 1 kbar (0.1 GPa). The highly converged results on forces are required for the calculation of the dynamical matrix using the direct force constant approach (or supercell method) **[44]**. The construction of the dynamical matrix at the Γ point is particularly simple and involves separate calculations of the forces in which a fixed displacement from the equilibrium configuration of the atoms within the *primitive* unit cell is considered. Symmetry aids by reducing the number of such independent distortions, reducing the amount of computational effort in the study of the analyzed structures considered in our work. Diagonalization of the dynamical matrix provides both the frequencies of the normal modes and their polarization vectors, it allows to us to identify the irreducible representation and the character of the phonon modes at the zone center.

**IV. Results and discussion**

**A. Raman measurements of the low-pressure phase**

A group-theoretical analysis of the wolframite structure of $ZnWO_4$ yields 36 lattice modes at the Γ point: 8 $A_g$ + 10 $B_g$ + 8 $A_u$ + 10 $B_u$, with the 18 even (g) vibrations being Raman active modes: 8 $A_g$ + 10 $B_g$. Symmetry assignments of the modes in the wolframite structure have been previously made for all the eighteen phonons by applying polarization selection rules in ZnWO$_4$ **[32, 45, 46]**. The symmetry assignments reported in **Ref. 46** are shown in **Table II** and have been confirmed by our *ab initio* calculations, as we will discuss latter. In scheelite-type AWO$_4$ compounds it has been proved that one can distinguish two types of vibrational modes: internal and external modes with respect to the WO$_4$ tetrahedra. The internal modes correspond to normal motions of atoms inside the WO$_4$ tetrahedra, while the external modes involve motions of WO$_4$ tetrahedra against the A atom. It is expected that the phonon frequencies of the internal modes will be higher that those of the external modes because the internal covalent bonding within the WO$_4$ tetrahedra is stronger than the external lattice binding. Due to the incompressibility of the WO$_4$ tetrahedra, it is also expected that the Grüneisen parameters of the internal modes were smaller than those of the external modes. Out of the internal modes, there are four stretching modes arising from each of the four W-O bonds in the WO$_4$ tetrahedra. A similar reasoning can be applied to AWO$_4$ wolframites assuming the incompressibility of WO$_6$ octahedra with respect to the ZnO$_6$ octahedra. Therefore, one would expect six internal stretching modes arising from each of the six W-O bonds in the WO$_6$ octahedra. These six internal stretching modes have been assigned in the literature by means of pressure and temperature dependent Raman studies and by comparison with other compounds **[32, 45, 46]**. However, there are important contradictions among different authors on this assignment.

**Figure 2** shows the RT Raman spectra of sanmartinite measured in sample #2 at selected pressures up to 40.2 GPa. The Raman spectra should correspond to a mixture of polarizations because of the use of a powder sample. In **Fig. 2**, it can be seen that up to 33.3 GPa it is possible to clearly distinguish the eighteen Raman modes of wolframite

ZnWO$_4$. **Table II** summarizes the phonon frequencies we measured at ambient pressure (0.0001 GPa) and compared them with those reported in the literature and those we calculated using *ab initio* lattice dynamics. The agreement between our results and those previously published **[32, 45, 46]** is quite good. **Table II** also summarizes the pressure coefficients (dω/dP) of the Raman modes of sanmartinite, and their Grüneisen parameters ($\gamma = B_0 / \omega \cdot d\omega / dP$, where B$_0$ is the bulk modulus). **Figure 3** shows the pressure evolution of the wolframite phonons extracted from the two different samples we studied. Results obtained from the two samples agree very well among themselves. From our experiments we determine slightly different pressure dependences for the Raman phonons than Perakis *et al.* **[32]**. In fact, for three phonons (A$_g$ mode near 342 cm$^{-1}$, B$_g$ mode near 313 cm$^{-1}$, and B$_g$ mode near 190 cm$^{-1}$) our pressure coefficients almost double the values previously reported (see **Table II**). At present we have no explanation for the observed differences in the pressure coefficients between the two high-pressure works. However, the fact that our *ab initio* calculations are in much better agreement with our measurements (see **Table II**) gives additional support to the accuracy of our measurements.

In order to calculate the mode Grüneisen parameters in wolframite ZnWO$_4$ its bulk modulus at zero pressure is needed. Since this magnitude has not been experimentally determined yet, we used the value we obtained from our *ab initio* calculations (B$_0$ = 140 GPa) to calculate the mode Grüneisen parameters reported in **Table II**. The bulk modulus of AWO$_4$ wolframite tungstates can be also estimated from the cation formal charge of the element A and the mean A-O distance using the empirical law reported in **Ref. 17**. This law was originally established for ABX$_4$ compounds with the scheelite, zircon or similar structures, in which the BX$_4$ tetrahedral units have a very low polyhedral compressibility. However, as a first approximation, the same law can also be applied to wolframite compounds since in these compounds the $WO_6$ octahedra are significantly less compressible than the $AO_6$ octahedra **[47]**. Within this framework, we obtain for $ZnWO_4$ $B_0 = 130 \pm 8$ GPa and for $CdWO_4$ $B_0 = 120 \pm 8$ GPa. The second value is in good agreement with the bulk modulus obtained from the low-pressure data reported by Macavei and Schulz **[47]**, confirming the predictive capability of the model developed by Errandonea *et al.* **[17]**. The obtained bulk modulus for $ZnWO_4$ (130 GPa) is close to our theoretical calculations (140 GPa).

In a first attempt to identify the six internal stretching modes of the W-O atoms in the distorted $WO_6$ octahedra of $ZnWO_4$, Liu *et al.* assigned them to the modes at 906, 787 and 407 $cm^{-1}$ on the basis of the bond lengths and Raman frequencies in the $WO_6$ group **[45]**. Afterwards, Wang *et al.* assigned the internal stretching modes to the phonons observed near 906 $cm^{-1}$, 787 $cm^{-1}$, 709 $cm^{-1}$, 407$cm^{-1}$, 342 $cm^{-1}$, and 190 $cm^{-1}$ on the basis of the temperature dependence of the Raman frequencies **[46]**. However, this assignment is in contradiction with the fact that the frequencies of the internal modes are expected to be higher than those of the external modes. These authors argue in favour of their assignment that the oxygen sharing between $WO_6$ and $ZnO_6$ octahedra may cause a considerable overlap in the frequency range for the two types of vibrations. This overlapping between internal and external modes was already discussed in scheelite tungstates and molybdates by Tarte and Liegeois-Duyckaerts, who claimed that no clear distinction can be made between internal and external modes **[48, 49]**. More recently, Perakis *et al.* **[32]** based upon high-pressure Raman-spectroscopy measurements corrected the previous assignment for the internal stretching modes including the phonon at 677 $cm^{-1}$ and not the phonon at 190 $cm^{-1}$. Our measurements and calculations support this correction, but also suggest that the $A_g$ phonon at 545 $cm^{-1}$ is an internal stretching mode in contrast to the $A_g$ phonon at 342 $cm^{-1}$ previously proposed. The new assignment of the internal stretching modes can be seen in **Table II**. It is worth commenting here, that after the assignment we proposed for the internal stretching modes, they consist of the four $A_g$ modes and the two $B_g$ modes with the highest frequencies, which is in fully agreement with the idea that in $AWO_4$ compounds the internal stretching modes are the highest in frequency **[19, 20]**. As we will discuss later this conclusion is also supported by our lattice-dynamics calculations.

Recently, the assignment of the internal stretching modes has been obtained in scheelite-structured orthotungstates after relating the frequencies (ω) of the stretching W-O modes inside the $WO_4$ tetrahedra with the Pauling's bond strengths (S) **[19, 20]**. Using the same approach of Hardcastle and Wachs for tungsten wolframites **[50]**, we can also obtain the Pauling's bond strengths in valence units (v.u.) from the stretching W-O mode frequencies (given in cm$^{-1}$), which for tungsten oxides is **[19, 20]**:

$$S_{W-O} = [0.27613 \ln (25823/\omega)]^{-6} \qquad (1)$$

Thus, it is possible to estimate the formal valence of the W ion in $ZnWO_4$ if we consider all the stretching frequencies of the internal modes of the $WO_6$ octahedra. By taking as these frequencies those we proposed (see **Table II**), and considering that the coordination of W in the wolframite structure is 4 + 2 = 6, we get the estimated total valence of 0.442 + 0.684 + 0.969 + 1.044 + 1.244 + 1.598 = 5.981 (in v.u.) in perfect agreement with the formal valence of the W ion. This fact gives additional support to the new assignment of the internal stretching modes we are proposing here. Following the same procedure, we can suggest that the internal stretching modes in wolframite $CdWO_4$ and $NiWO_4$ are at: 900, 775, 710, 690, 550, and 390 cm$^{-1}$ ($CdWO_4$), and 890, 775, 691, 678, 556, and 424 cm$^{-1}$ ($NiWO_4$).

**B. Raman measurements of the high-pressure phase**

In **Fig. 2**, it is possible to see that some changes take place in the Raman spectra from 30.6 to 40.2 GPa. First we observed the appearance of eight new peaks (depicted by ticks in **Fig. 2**) in addition to the eighteen wolframite peaks at 30.6 GPa. In particular, the new peak located around 900 cm$^{-1}$ is quite strong. At 33.3 GPa new peaks emerge, reaching the total number of fourteen, while the wolframite peaks can still be clearly observed. It also becomes clear that the strong and broad new peak, located around 900 cm$^{-1}$, consists of a triplet. At 36.5 GPa most of the wolframite peaks become very weak and only some of them (e.g. the strongest peak of wolframite located at 1030 cm$^{-1}$ at this pressure) can be observed. At this pressure the new peaks are already sixteen. At 40.2 GPa, all the wolframite peaks have disappeared and only the sixteen new peaks are present. **Table III** gives the frequencies of the new peaks observed at 40.2 GPa and their pressure coefficients. We think that the changes in the Raman spectra are caused by the occurrence of a pressure-induced phase transition. The onset of the transition is located at 30.6 GPa, and a large coexistence region, of both the low- and high-pressure phases, is observed between 30.6 and 36.5 GPa. The phase transition is fully completed at 40.2 GPa and it is fully reversible with very little hysteresis, as can be seen in the spectrum collected at 0.7 GPa after pressure release. In this spectrum all the observed peaks can be assigned to the eighteen Raman modes of wolframite.

Very similar changes to those we observed in ZnWO$_4$ from 30.6 GPa to 40.2 GPa were found by Jayaraman *et al.* in CdWO$_4$ between 20 and 28 GPa **[31]**. In CdWO$_4$, also sixteen modes can be observed in the high-pressure phase, and they are located at similar frequencies as those we detected for the high-pressure phase of ZnWO$_4$. The phonons for the high-pressure phase of both compounds can be compared in **Table III**. It can be seen that the phonons of CdWO$_4$ resemble those of ZnWO$_4$, but shifted to lower frequencies, in both the high-pressure and low-pressure phases, due to the larger atomic mass of the Cd cation (see **Tables II and III** and **Ref. 31**). A similar mass-dependent shift is observed in the Raman spectra of alkaline-earth tungstates following the series Ba, Sr, Ca **[19, 20]**. The similitude of the Raman spectra reported by us for $ZnWO_4$ at 40.2 GPa and that reported by Jayaraman for $CdWO_4$ at 35 GPa suggests that the structure of the high-pressure phases of $ZnWO_4$ and $CdWO_4$ can be the same, in a similar way to what is observed in scheelite tungstates **[18 - 25]**. A remarkable feature is that the $A_g$ mode that represents the totally symmetric W-O stretching vibration ($\omega = 907$ cm$^{-1}$ at ambient pressure and $\omega = 1013$ cm$^{-1}$ at 30.6 GPa) in wolframite $ZnWO_4$ drops by about 110 cm$^{-1}$ in $ZnWO_4$ and 120 cm$^{-1}$ in $CdWO_4$ at the phase transition (see **Fig. 3** and **Ref. 31**). This fact suggests that a W-O coordination increase takes place at the phase transition. When the coordination of W increases, the W-O bond lengths usually increase too, and the result is a drop in the frequency of the internal stretching modes. This kind of behavior is observed for instance in $Al_2(WO_4)_3$ **[51]** and in the scheelite tungstates **[19, 20]** after a pressure-induced phase transition that imply an increase of the W-O coordination from tetrahedral to octahedral. Another interesting feature observed in the Raman spectra of the high-pressure phases of $ZnWO_4$ and $CdWO_4$ is the appearance of new modes in the phonon-gap region of their low-pressure phases (e. g. between 470 and 600 cm$^{-1}$ in ZnWO4 at 30 GPa, see Fig. 3; for $CdWO_4$ see Fig. 2 of **Ref. 31**). These two facts will be very helpful to identify the structure of the high-pressure phase of $ZnWO_4$.

Before closing this subsection, we would like to mention that a formation of domains has been observed around 12 GPa in the single crystalline sample (#1). This domain formation occurs together with a relative change of the phonon intensity. In particular, the $B_g$ modes, which were much weaker than the $A_g$ modes because of polarization selection rules [52] below 12 GPa, gain in intensity. A similar behavior was described for single crystalline $CdWO_4$ near 10 GPa [31] and was interpreted as a phase transition, even though it did not cause any evident change in the Raman spectra or in the pressure evolution of the phonons. The attribution of the domain formation to structural phase transitions is challenged by the fact that in powder $ZnWO_4$, Perakis *et al.* [32] did not observe any phase transition up to 24 GPa and we did not observe the onset of it up to 30.6 GPa. A different and more suitable interpretation of this phenomenology could be based on the the formation of permanent defects on the sample, as observed recently in optical-absorption measurements in $ZnWO_4$ near 12 GPa [40]. The presence of these defects will cause a breaking of the local structural symmetry producing the increase of the intensity of the $B_g$ modes. The fact that this increase is gradual, and simultaneous with the domain formation, suggests that the domains, observed by us in the single crystal of $ZnWO_4$ and by Jayaraman *et al.* in single crystals of $CdWO_4$, are the consequence of the gradual introduction of defects, which can be precursors of the phase transition observed at much higher pressures [53]. Indeed, the introduction of such defects, which leads to a structure with domains separated by antiphase boundaries was observed in wolframite-structured $FeNbO_4$ [54].

Finally, it is well-known that in scheelite-structured $ABX_4$ compounds, the pressure at which the low-pressure phase becomes unstable, is correlated with the $BX_4/A$ radii ratio; the larger this ratio, the higher the transition pressure [30]. Applying the relationship proposed in **Ref. 30** to wolframite-structured $ZnWO_4$ and $CdWO_4$ we estimate that a pressure-induced phase transition should be expected somewhere beyond 24 GPa and 18 GPa, respectively. These pressures, are close to the pressure were the onset of the phase transition is observed in the Raman experiments. This fact supports that only one phase transition is observed in $ZnWO_4$ and $CdWO_4$, being the onset of the transition at 30.6 GPa and 20 GPa, respectively, and the transition completed at 40.2 GPa and 30 GPa, respectively.

## C. *Ab initio* calculations

In a recent work, Manjón *et al.* have shown experimentally with the help of *ab initio* calculations that pressure-induced phase transitions in $ABX_4$ compounds, in particular in scheelite tungstates and zircons, follow the phenomenological North-East rule in Bastide's diagram **[23]**. In this sense, we present now the results from our theoretical total-energy calculations of several structural phases of $ZnWO_4$ in order to study the structural stability of the wolframite structure and its possible high-pressure phases. Along with the wolframite structure (SG: *P2/c*) **[10]**, we have considered other structures, paying special attention to phases located to the North-East with respect to wolframite in Bastide's diagram, on account of their observation or postulation in previous high-pressure works for related compounds: $CuWO_4$-type (SG: $P\bar{1}$) **[55]**, orthorhombic-disorder wolframite (SG: P*bcn*) **[56]**, M-fergusonite (SG: *I2/a*) **[57]**, M'-fergusonite (SG: *P2$_1$/c*) **[58]**, $YNbO_4$-type β-fergusonite (SG: *C2/c*) **[59]**, monoclinic-distorted rutile (SG: *P2/c*) **[60]**, scheelite (SG: *I4$_1$/a*) **[61]**, $HgWO_4$-type (SG: *C2/c*) **[62]**, *Cmca* (SG: *Cmca*) **[18]**, BaWO4-II-type (SG: *P2$_1$/n*) **[63]**, baddeleyite (SG: *P2$_1$/c*) **[64]**, and α-$SnWO_4$ (SG: *Pnna*) **[65]**. **Figure 4** shows the energy-volume curves for the different structures of $ZnWO_4$ from which the relative stability and coexistence pressures of the phases can be extracted by the common-tangent construction **[66]**. In this figure, we only reported those structures that play a relevant role in the high-pressure structural behavior of $ZnWO_4$. From **Figure 4** it is deduced that the wolframite phase is stable at zero and low pressure up to 39 GPa, with an equation of state (EOS) with parameters $V_0$= 137.4 Å$^3$, $B_0$= 140 GPa, and $B_0$'= 4.57, where the parameters $V_0$, $B_0$, and $B_0$' are the zero-pressure volume, bulk modulus, and pressure derivative of the bulk modulus, respectively. These parameters were obtained from our calculations using a third-order Birch-Murnaghan EOS **[67]**. In the inset of **Fig. 4** we show the calculated P-V relationship obtained for wolframite $ZnWO_4$, which corresponds to the above reported EOS. The obtained lattice parameters at ambient pressure compare well with the experimental results with differences within the typical reported systematic errors in DFT-GGA calculations (see **Table I**). A similar degree of agreement exists for the calculated values of the internal parameters of the wolframite phase.

At low pressures there are three structures in close competition with the wolframite structure: the triclinic $P\bar{1}$, the baddeleyite, and the distorted-rutile structures. The triclinic $CuWO_4$-type structure, is known to be a metastable phase in wolframite-type $ZnMoO_4$ **[68]**. In $ZnWO_4$, the $CuWO_4$-type structure is expected to become stable at expanded volumes (i.e. *negative* pressures) according to the phase-transition systematics established by Bastide **[23, 26]** based on the ionic radio ratios of the A and B cation and the X anion in $ABX_4$ compounds. Therefore, it is not strange that this structure is competitive with wolframite. As a matter of fact, the *P2/c*-to-$P\bar{1}$ transition is observed in solid solutions of sanmartinite and cuproscheelite ($CuWO_4$) at around $Zn_{0.78}Cu_{0.22}WO_4$ **[69]**. This fact suggests that the $P\bar{1}$-to-*P2/c* phase transition could be observed in $CuWO_4$ upon compression. This picture is consistent with the idea that under pressure the electronic structure of an element of the Periodic Table becomes similar to that of the next-row element as a consequence of the pressure-induced *sp-d* electron transfer **[70, 71]**. In addition, the group-subgroup relationship existent between the $P\bar{1}$ and *P2/c* space groups makes the proposed transition quite reasonable from the crystallochemical point of view **[72]**. On the other hand, the baddeleyite and distorted-rutile structures have been also observed as metastable phases in compounds isostructural to ZnWO$_4$, e.g. in FeNbO$_4$ **[60]**. Therefore, it is not strange that these structures are in close competition with the wolframite structure at low pressures.

**Figure 5** shows the pressure dependence we obtained from our calculations for the lattice parameters and the monoclinic β angle of wolframite ZnWO$_4$. There, it can be seen that as in other orthotungstates the compressibility of ZnWO$_4$ is highly anisotropic. In particular the *c*-axis is much less compressible than the other two axes. For example, from ambient pressure to 20 GPa the relative compression of the *a*- and *b*-axis is approximately 4%, but the relative compression of the *c*-axis is only 0.8%. Consequently, the *b/c* axial ratio decreases considerably upon compression whereas the *b/a* axial ratio stays nearly constant. A similar behavior has been observed in CdWO$_4$ up to 8 GPa **[47]**. At the same time, compression also causes a reduction of the β angle. On the other hand, the changes of the atomic positions with pressure are negligible (see **Table I**). The anisotropic compressibility of wolframite ZnWO$_4$ can be understood in terms of hard anion-like WO$_6$ octahedra surrounded by charge compensating Zn cations. **Figure 6** shows the calculated pressure evolution for the Zn-O and W-O interatomic distances. As can be seen there, there is an important decrease upon compression of the Zn-O bond distances, but the W-O bond distances are nearly uncompressible. In particular, the larger Zn-O bonds are the most compressible bonds, which cause a gradual reduction of the anisotropy of the ZnO$_6$ octahedra upon compression. A similar uncompressibility of the W-O bonds has been observed in the case of the scheelite-structured orthotungstates **[17, 18]**. This means that when pressure is applied the WO$_6$ units remain essentially undistorted and the reduction of the unit-cell size is basically associated to the compression of the Zn-O octahedral environment. Along the *c*-axis the WO$_6$ units are directly aligned, whereas along the *a*- and *b*-axis there is a Zn cation between two WO$_6$ octahedra. Thus, the different arrangement of hard WO$_6$ octahedra along the *a*-, *b*-, and *c*-axis accounts for the different compressibility of the three unit-cell axes. The uncompressibility of the $WO_6$ octahedra in wolframite-structured orthotungstates, which is similar that of $WO_4$ tetrahedra in scheeliite-structured orthotungstates, explains why the empirical relation proposed in **Ref. 17** for the bulk modulus of scheelite-structured $AWO_4$ compounds also works accurately for wolframite $ZnWO_4$ and $CdWO_4$.

As pressure increases, the wolframite structure becomes unstable against a monoclinic β-fergusonite-type structure. Theoretically, this structure only emerges as a structurally different and thermodynamically stable phase above a compression threshold of about 39 GPa. At the transition pressure the atomic volume of the wolframite phase is 114.89 $Å^3$ (two formula units per unit cell) and the atomic volume of the β-fergusonite phase is 215.54 $Å^3$ (four formula unit per unit cell). Thus the occurrence of the phase transition implies a large volume collapse of about 6%, which suggest that the transition is a first-order reconstructive transformation. The structural parameters obtained for the β-fergusonite structure at 44.1 GPa are given in **Table IV** and a perspective drawing of it is shown in **Figure 1(b)**. As can be seen in the figure, in the high-pressure phase the packing is more compact than in the wolframite structure, being in the β-fergusonite-type structure the W atoms coordinated by four O atoms at 1.84 Å and four additional oxygens at 2.64 Å. On the other hand, in the high-pressure phase the Zn atoms are coordinated by 4 four atoms at a short distance of about 1.95 Å and by four O atoms at a longer distance of about 2.04 Å, forming a distorted dodecahedra. This fact implies an increase of the W-O coordination from 4 + 2 to 4 + 4, which is in good agreement with the drop of the W-O stretching mode observed at the phase transition. According with our calculation the monoclinic β-fergusonite phase of $ZnWO_4$ remains as the most stable phase up to 57.6 GPa. On further increase of pressure we found that the orthorhombic *Cmca* structure that we proposed in a previous study for CaWO$_4$ and SrWO$_4$ **[18]** becomes favoured beyond 57.6 GPa. Unfortunately, our experiments could not be extended up to this pressure in order to check our theoretical prediction. The crystal parameters of the second high-pressure phase are given in **Table V** and a perspective drawing of it is shown in **Figure 1(c)**. The most interesting feature of this phase is that it implies a W-O coordination increase from 4 + 4 to 8 and a Zn-O coordination increase from 8 to 7 + 4.

The high-pressure structural sequence we are reporting here (wolframite → β-fergusonite → *Cmca*) can be rationalized by means of the phase diagram proposed by Bastide **[23, 26]**. In this diagram the ABX$_4$ compounds are located according with their cation-to-anion radii ratios ($r_A/r_X$, $r_B/r_X$) and expected to undergo pressure-induced phase transitions following the north-east rule; i.e. a given compound is expected to take the structure of a compound with larger cation-to-anion radii ratios. According with this picture, ZnWO$_4$ could probably transform under pressure to the structure of YNbO$_4$ (β-fergusonite) and after that to an orthorhombic *Cmca* structure similar to that of BaMnF$_4$ and SrUO$_4$. Please note that it is possible to transform wolframite into β-fergusonite by means of a *klassengleiche* transformation and this structure into *Cmca* by means of *translationengleiche* transformation. Therefore, given the group relationships existent among the three structures, the structural sequence we are proposing here for ZnWO$_4$ is definite possible on crystallographic grounds. It is important to mention here that the high-pressure β-fergusonite phase of ZnWO$_4$ is closely related from a structural point of view to the high-pressure M-fergusonite phase found in scheelite-structured AWO$_4$ compounds **[17, 18]**. β-fergusonite can be obtained by means of symmetry operations from wolframite and M-fergusonite can be obtained by means of symmetry operations from scheelite **[73]**. Both fergusonite structures are not isostructural but are closely related because both structures consist of zigzag chains of W polyhedra with eight coordinated A atoms. Apparently, both in wolframite- and scheelite-structured compounds the ferguosonite phases act as a bridge phase between a structure with a low W coordination (like scheelite and wolframite) and another with high W coordination, like the *Cmca* structure.

**Table II** shows the calculated frequencies and pressure coefficients for the Raman modes of wolframite $ZnWO_4$. The agreement between calculations and experiments for the low-pressure phase is quite good, which gives credibility to the lattice-dynamics calculations we performed for the high-pressure phase of $ZnWO_4$. It is important to note here that the calculated eigenmodes for wolframite $ZnWO_4$ indicate that there are modes which involve basically a movement of the $WO_6$ octahedra as rigid units and other that imply internal vibrations of these octahedra. Therefore, despite external and internal modes show similar Grüneisen parameters and this prevents a simple distinction between internal and external modes, as in scheelite-structured compounds, this distinction can be still applied to wolframite-structured compounds as we did along the paper. The isolation of the $WO_6$ octahedra is also evident from the uncompressibility of the W-O already described. According with our lattice-dynamics calculations the internal stretching modes are the same we proposed in Section II.A.

Let us know discuss the lattice-dynamic calculations performed for the high-pressure phase of $ZnWO_4$. The present Raman measurements are in good agreement with the results obtained from our calculations beyond 39 GPa. In particular, the Raman spectra collected for the first high-pressure phase are best explained for the β-ferguosonite structure. Other candidate structure cannot give account of the measured Raman spectra. In the first place, the scheelite structure in tungstates has only thirteen Raman active modes and has a phonon gap from 400 $cm^{-1}$ to 700 $cm^{-1}$ **[19, 20]**. So it cannot explain the Raman spectra we observed in the high-pressure phase and therefore is discarded as a candidate structure for the high-pressure phase. In the second place, the M-fergusonite and M'-fergusonite structures in tungstates and the $HgWO_4$ structure have eighteen Raman active modes [74], but they have their strongest internal stretching mode well beyond 900 cm$^{-1}$; i.e., as high as those in the scheelite and wolframite phases [19, 20]. So these structures cannot explain the drop we observed in the stretching mode basically because they do not imply an increase in the W-O coordination. In addition, these structures usually also have a phonon gap, which is not present in our Raman spectra. Finally, we have found that the β-fergusonite structure gives a phonon spectrum that can explain reasonably well our experimental results. According to group-theoretical considerations the β-fergusonite has 18 Raman active modes at the Γ point: 8 $A_g$ + 10 $B_g$. The frequencies and mode assignment of the different phonon calculated for this structure at 40 GPa are given in **Table III**. According with the calculations there are always two $B_g$ modes very close in frequency to each other, but lattice-dynamics calculations apparently tend to underestimate the frequency splitting between $B_g$ modes in fergusonite structures [19, 20]. However, the calculated small splitting between $B_g$ modes could explain why in the experiments we have only found sixteen modes, since some of the $B_g$ modes could be degenerated within the accuracy of the experiments. The qualitative agreement between the calculated and measured phonon frequencies and pressure coefficients is reasonably good. Indeed, among the different structures considered in the calculation, the β-fergusonite structure is the only one that gives a good quantitative agreement with the experiments. In addition, Raman spectra measured in the β-fergusonite phase of $YNbO_4$ [75] resemble very much those measured for the high-pressure phases of $ZnWO_4$ and $CdWO_4$. Therefore, we conclude that we found enough evidence to propose that the high-pressure phase of wolframite-structured AWO$_4$ compounds have a distorted β-fergusonite structure.

## V. Conclusions

We have performed RT Raman scattering measurements under pressure in ZnWO$_4$ up to 45 GPa. The frequency pressure dependence of all first-order modes of the wolframite phase have been measured up to the completion of the scheelite-to-β-fergusonite phase transition around 40 GPa. This value of the transition pressure is in good agreement with the estimated transition pressure (39 GPa) according to our *ab initio* total-energy calculations. Our measurements show that the transition to the β-fergusonite phase starts at 30.6 GPa but it is not completed up to 40.2 GPa. The phase transition is reversible and occurs with a volume collapse of about 6%. The *ab initio* calculations also allows us to determine the pressure evolution of the unit-cell parameters of wolframite ZnWO$_4$, being observed that its compression is highly anisotropic. This behavior is related to the different compressibility of Zn-O and W-O bonds, being the last ones much more rigid than the first ones. The calculations also suggest the occurrence of a second pressure-induced phase transition from the β-fergusonite structure to an orthorhombic *Cmca* structure. Additionally, we have performed *ab initio* lattice dynamics calculations for ZnWO$_4$ at selected pressures in the wolframite and β-fergusonite phases. Our calculated mode frequencies in both structures agree with the frequencies of the observed Raman modes and have allowed the assignment and discussion of the nature of the modes.

## Acknowledgments

The authors thank A. Cantarero (ICMUV, Universidad de Valencia) for providing us access to the experimental Raman setup. This work was made possible through financial support of the MCYT of Spain under grants No. MAT2007-65990-

C03-01, MAT2007-65990-C03-03, No. MAT2006-02279, and CSD2007-00045, of the Generalitat Valenciana under grant No. ACOMP06/81 and GV06/151, and of the Nature Science Foundation of the Fujian Province of China under grant No. 2005HZ1026. D.E. acknowledges the financial support from the MCYT of Spain through the "Ramon y Cajal" program. F.J.M. acknowledges the financial support from "Vicerrectorado de Innovación y Desarrollo de la UPV" through project UPV2008-0020. T.Y. also acknowledges the support received through the great project of FJIRSM (SZD08001-2).

**Table I:** Experimental and calculated crystal parameters of wolframite ZnWO$_4$. Space group *P2/c*, Z = 2. Calculated parameters are given at different pressures.

|  | Neutron Diffraction (**Ref. 34**) Ambient Pressure | X-ray Diffraction (This Work) Ambient Pressure | *Ab initio* Calculations (This Work) Ambient Pressure | *Ab initio* Calculations (This Work) P = 27.2 GPa |
|---|---|---|---|---|
| *a* | 4.693 Å | 4.680 Å | 4.741Å | 4.516Å |
| *b* | 5.721 Å | 5.712 Å | 5.824 Å | 5.521 Å |
| *c* | 4.928 Å | 4.933 Å | 4.977 Å | 4.799 Å |
| *β* | 90.632º | 90.3º | 90.759º | 89.899º |
| Zn Site: 2f | x = 0.5<br>y = 0.6833<br>z = 0.25 | x = 0.5<br>y = 0.697<br>z = 0.25 | x = 0.5<br>y = 0.6811<br>z = 0.25 | x = 0.5<br>y = 0.6780<br>z = 0.25 |
| W Site: 2e | x = 0<br>y = 0.1823<br>z = 0.25 | x = 0<br>y = 0.178<br>z = 0.25 | x = 0<br>y = 0.1813<br>z = 0.25 | x = 0<br>y = 0.1873<br>z = 0.25 |
| O$_1$ Site: 4g | x = 0.2547<br>y = 0.3772<br>z = 0.4005 | x = 0.244<br>y = 0.372<br>z = 0.394 | x = 0.2561<br>y = 0.3741<br>z = 0.4025 | x = 0.2570<br>y = 0.3908<br>z = 0.4101 |
| O$_2$ Site: 4g | x = 0.2171<br>y = 0.8955<br>z = 0.4360 | x = 0.203<br>y = 0.904<br>z = 0.456 | x = 0.2153<br>y = 0.8943<br>z = 0.4365 | x = 0.2292<br>y = 0.9009<br>z = 0.4348 |

**Table II:** *Ab initio* calculated and experimental zero-pressure frequencies, pressure coefficients, and Grüneisen parameters of the Raman modes in wolframite ZnWO$_4$. The asterisks indicate the internal stretching modes. The Grüneisen parameter has been calculated using the calculated bulk modulus B$_0$= 140 GPa as indicated in the text.

| Mode | Experiment[a] | | | Experiment[b] | | Theory[a] | |
|---|---|---|---|---|---|---|---|
| | ω [cm$^{-1}$] | dω/dP [cm$^{-1}$/GPa] | γ | ω [cm$^{-1}$] | dω/dP [cm$^{-1}$/GPa] | ω [cm$^{-1}$] | dω/dP [cm$^{-1}$/GPa] |
| B$_g$ | 91.5 | 0.95 | 1.45 | 91 | 1.3 | 83.7 | 1.02 |
| A$_g$ | 123.1 | 0.65 | 0.74 | 123 | 1.1 | 118.6 | 0.48 |
| B$_g$ | 145.8 | 1.2 | 1.15 | 145.5 | 2.05 | 137.2 | 1.33 |
| B$_g$ | 164.1 | 0.72 | 0.61 | 163.5 | 0.85 | 163.3 | 0.42 |
| B$_g$ | 189.6 | 0.67 | 0.49 | 189.5 | 0.32 | 182.2 | 0.41 |
| A$_g$ | 196.1 | 2.25 | 1.61 | 195 | 3.3 | 185.7 | 2.52 |
| B$_g$ | 267.1 | 1.32 | 0.69 | 266 | 1.25 | 261.2 | 2.16 |
| A$_g$ | 276.1 | 0.87 | 0.44 | 274 | 0.88 | 263.7 | 0.82 |
| B$_g$ | 313.1 | 1.74 | 0.78 | 314.5 | 1 | 298.3 | 1.44 |
| A$_g$ | 342.1 | 1.74 | 0.71 | 341.5 | 0.85 | 324.2 | 1.7 |
| B$_g$ | 354.1 | 3.87 | 1.53 | 355 | 4.6 | 342.1 | 3.3 |
| A$_g$ * | 407 | 1.65 | 0.57 | 407.5 | 1.4 | 383.8 | 1.84 |
| B$_g$ | 514.5 | 3.18 | 0.86 | 515.5 | 3.3 | 481.1 | 3.1 |
| A$_g$ * | 545.5 | 3 | 0.77 | 545 | 3.4 | 515.4 | 3.07 |
| B$_g$ * | 677.8 | 3.9 | 0.80 | 677 | 3.9 | 635.5 | 3.9 |
| A$_g$ * | 708.9 | 3.3 | 0.65 | 708.5 | 3.3 | 678.5 | 3.24 |
| B$_g$ * | 786.1 | 4.4 | 0.78 | 787 | 4.8 | 753.3 | 4.0 |
| A$_g$ * | 906.9 | 3.7 | 0.57 | 906 | 4.1 | 861.8 | 3.36 |

[a]This work, [b]**Ref. 32**.

**Table III:** Frequencies at 40 GPa and pressure coefficients of the Raman modes of the high-pressure phase of $ZnWO_4$. The frequencies and pressure-coefficients obtained after *ab initio* calculations are also given. For comparison, the same data form the high-pressure phase of $CdWO_4$ is given at 35 GPa [31].

| Mode | $ZnWO_4$ (40.2 GPa) Raman | | $ZnWO_4$ (40 GPa) Theory | | $CdWO_4$ (35 GPa) Raman | |
|---|---|---|---|---|---|---|
| | $\omega$ [cm$^{-1}$] | $d\omega/dP$ [cm$^{-1}$/GPa] | $\omega$ [cm$^{-1}$] | $d\omega/dP$ [cm$^{-1}$/GPa] | $\omega$ [cm$^{-1}$] | $d\omega/dP$ [cm$^{-1}$/GPa] |
| $A_g$ | 140.1 | 0.29 | 141.2 | 0.04 | 87 | 1.4 |
| $B_g$ | 179.9 | 0.72 | 184.6 | 0.09 | 112 | 0.4 |
| $B_g$ | | | 184.7 | 0.09 | 150 | 0.3 |
| $A_g$ | 208.1 | 1.30 | 226.2 | -0.19 | 183 | 0.4 |
| $B_g$ | 263.2 | 0.79 | 243.4 | 0.32 | 213 | 0.3 |
| $B_g$ | | | 243.5 | 0.32 | 240 | 0.3 |
| $A_g$ | 300.1 | 1.02 | 293.6 | 0.93 | 283 | 0.5 |
| $B_g$ | 339.1 | 1.24 | 300.3 | 0.70 | 322 | 1.7 |
| $B_g$ | 356.1 | 0.89 | 300.4 | 0.70 | 378 | 1.1 |
| $A_g$ | 402.1 | 2.07 | 375.8 | 0.97 | 439 | 2.7 |
| $A_g$ | 527.7 | 1.86 | 468.6 | 2.70 | 495 | 0.8 |
| $B_g$ | 588.5 | 2.54 | 591.4 | 2.97 | 572 | 1.2 |
| $B_g$ | 600.9 | 2.50 | 591.5 | 2.97 | | |
| $A_g$ | 752.7 | 2.56 | 715.3 | 2.74 | 677 | 1.7 |
| $A_g$ | 839.1 | 2.92 | 832.1 | 2.47 | 736 | 2.0 |
| $B_g$ | 886.1 | 2.30 | 898.1 | 2.73 | 768 | 1.4 |
| $B_g$ | 902.1 | 2.30 | 898.4 | 2.75 | | |
| $A_g$ | 928.0 | 3.32 | 906.9 | 2.54 | 867 | 1.7 |

**Table IV:** Calculated crystal parameters of the β-ferguson ite phase of ZnWO$_4$ at 44.1 GPa. Space group *C2/c*, Z = 4.

| a = 6.814 Å, b = 9.177 Å, c = 4.819 Å, and β = 134.976º | | | | |
|---|---|---|---|---|
| Atom | Site | x | y | z |
| Zn | 4e | 0 | 0.3750 | 0.25 |
| W | 4e | 0 | 0.8753 | 0.25 |
| O$_1$ | 8f | 0.1787 | 0.7994 | 0.1222 |
| O$_2$ | 8f | 0.3066 | 0.9507 | 0.7355 |

**Table V:** Calculated crystal parameters of the *Cmca* phase of ZnWO$_4$ at 76.1 GPa. Space group *Cmca*, Z = 8.

| a = 7.1807 Å, b = 10.3304 Å, and c = 4.9896 Å | | | | |
|---|---|---|---|---|
| Atom | Site | x | y | z |
| Zn | 8e | 0.75 | 0.8585 | 0.75 |
| W | 8f | 0.5 | 0.3920 | 0.2590 |
| O$_1$ | 8e | 0.75 | 0.1741 | 0.75 |
| O$_2$ | 8f | 0.5 | 0.2758 | 0.5444 |
| O$_3$ | 8d | 0.6600 | 0 | 0 |
| O$_4$ | 8f | 0.5 | 0.4217 | 0.8751 |

**Figure Captions**

**Figure 1:** (a) Perspective drawing of the crystal structure of wolframite $ZnWO_4$. (b) Perspective drawing of the crystal structure of the proposed β-fergusonite phase of $ZnWO_4$. (c) Perspective drawing of the crystal structure of proposed *Cmca* phase of $ZnWO_4$. Large circles: Zn, medium circles: W, and small circles: O. The conventional unit cell is represented with solid lines. W-O and Zn-O bonds are also shown as well as the different polyhedra.

**Figure 2:** Raman spectra of wolframite $ZnWO_4$ at different pressures. Ticks indicate the position of the Raman peaks assigned to the high-pressure phase. All the spectra were measured on pressure increase with the exception of the spectra marked with (r) which was taken after pressure release.

**Figure 3:** Pressure dependence of the Raman mode frequencies of the wolframite (solid symbols) and β−fergusonite (empty symbols) phases of $ZnWO_4$. Sample #1: squares. Sample #2: circles. The solid lines are just a guide to the eye. The vertical dashed lines indicate the range of coexistence of the wolframite and β−fergusonite phases.

**Figure 4:** Energy-volume curves calculated for $ZnWO_4$. Empty squares: wolframite struture, solid circles: $CuWO_4$-type structure, solid squares: β-fergusonite structure, and empty circles: *Cmca* structure. The inset shows the calculated pressure dependence of the volume (symbols) and the obtained equation of state (line) for the wolframite phase.

**Figure 5:** Theoretically-calculated pressure dependence of the lattice parameters and the monoclinic β angle of the wolframite structure of $ZnWO_4$.

**Figure 6:** Theoretically-calculated pressure evolution of the Zn-O and W-O interatomic distances in the low-pressure phase of $ZnWO_4$.

**Figure 1**

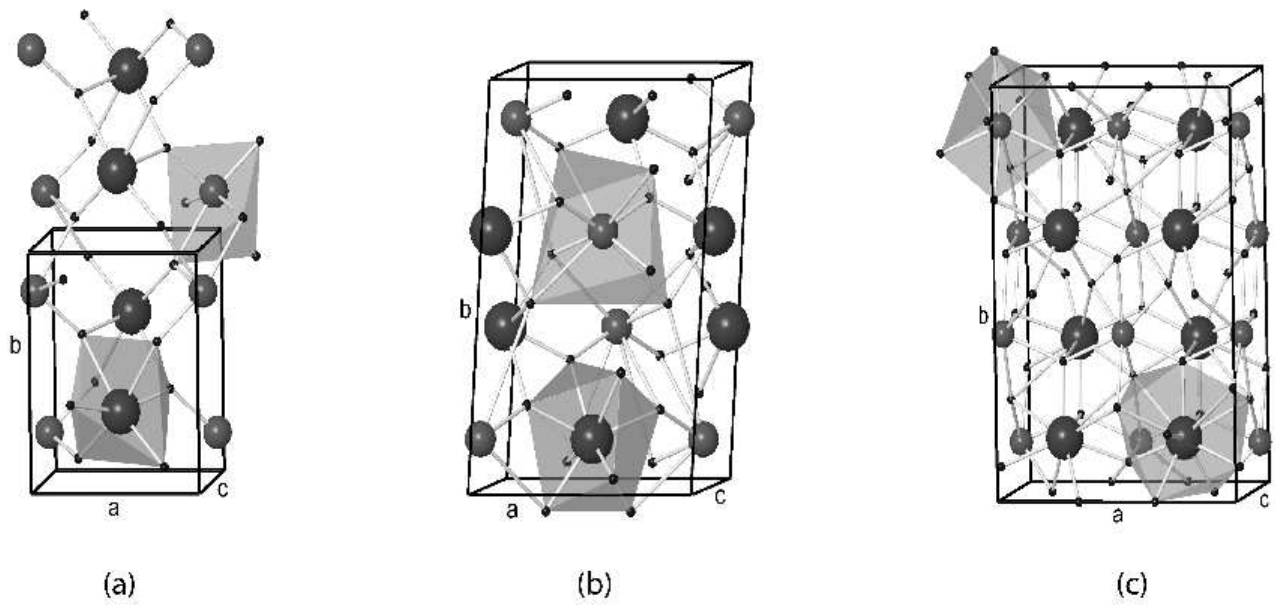

(a)          (b)          (c)

**Figure 2**

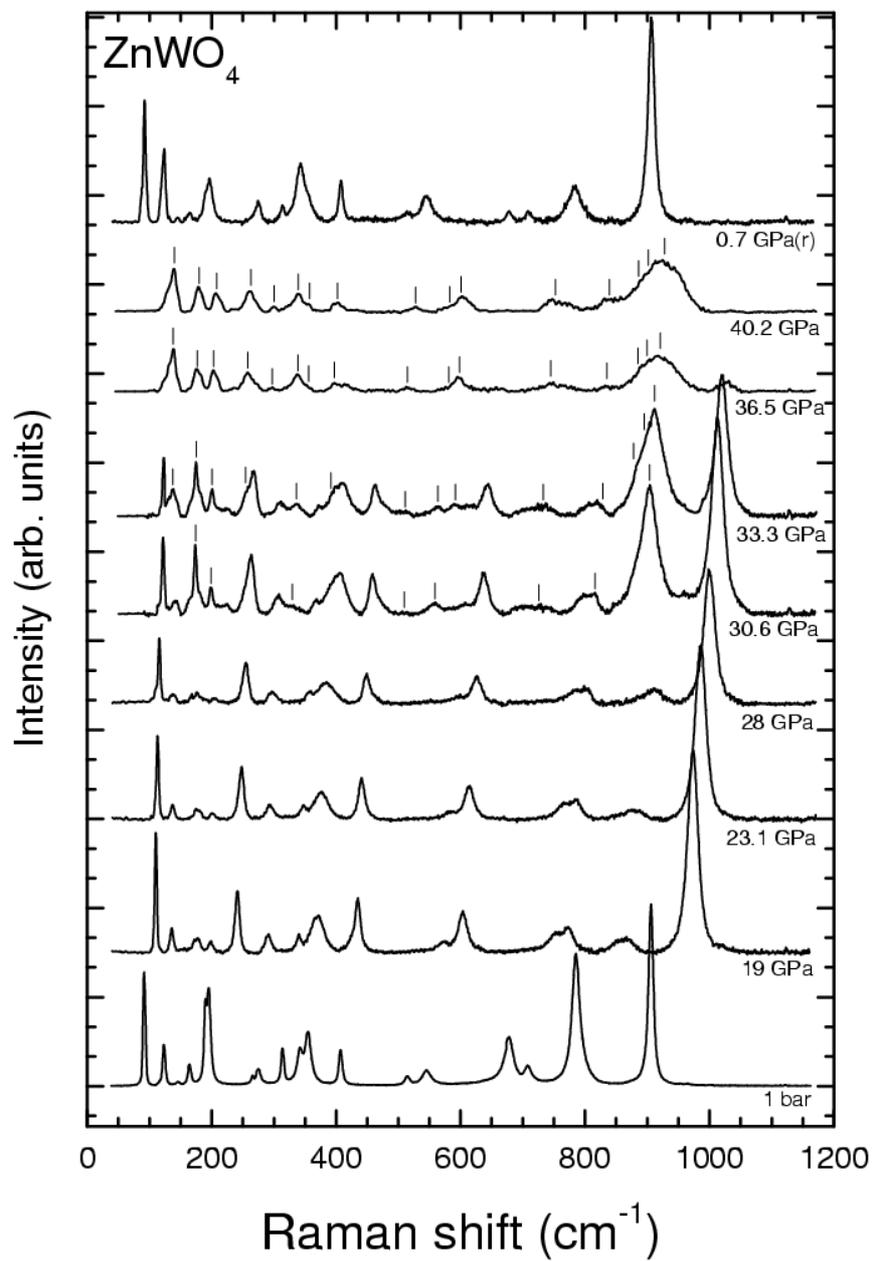

**Figure 3**

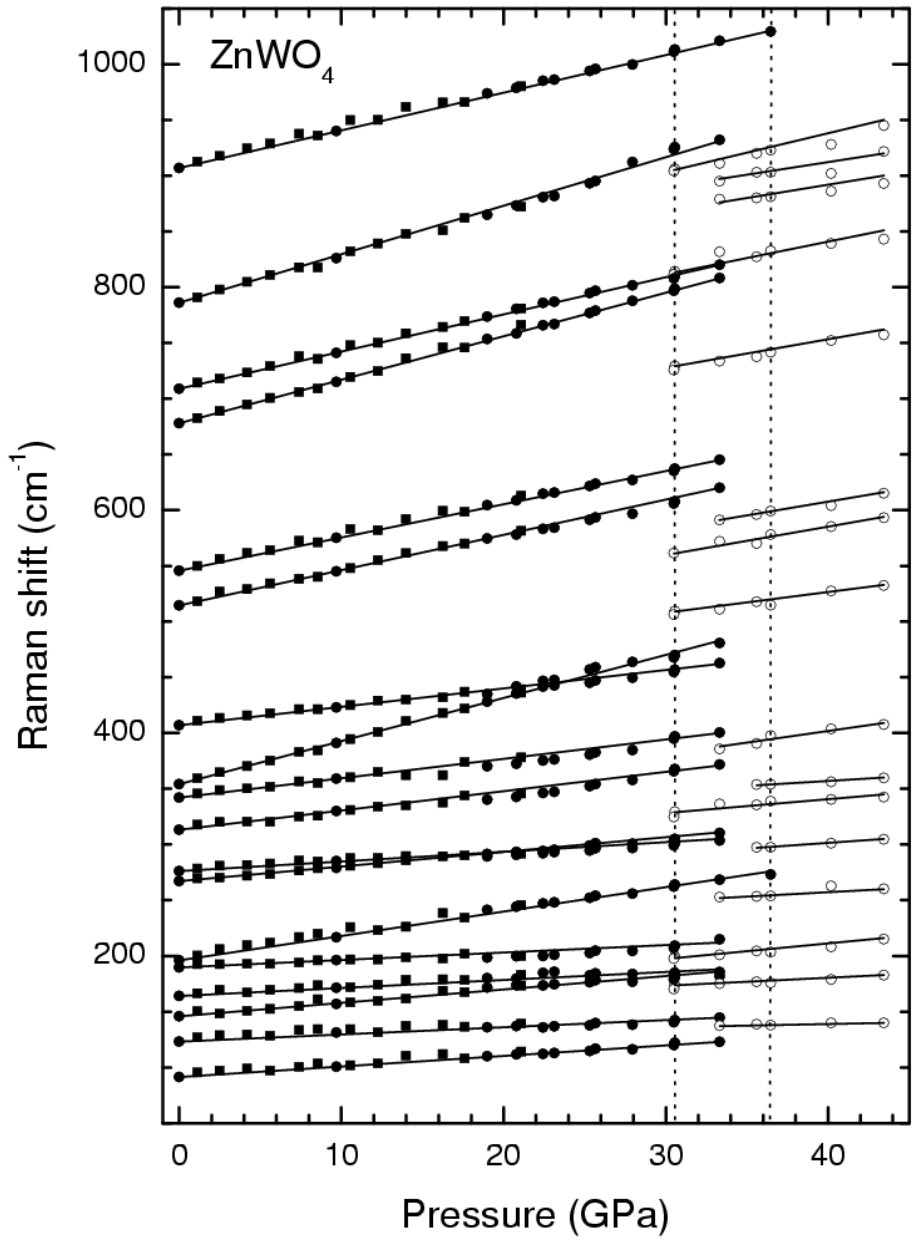

**Figure 4**

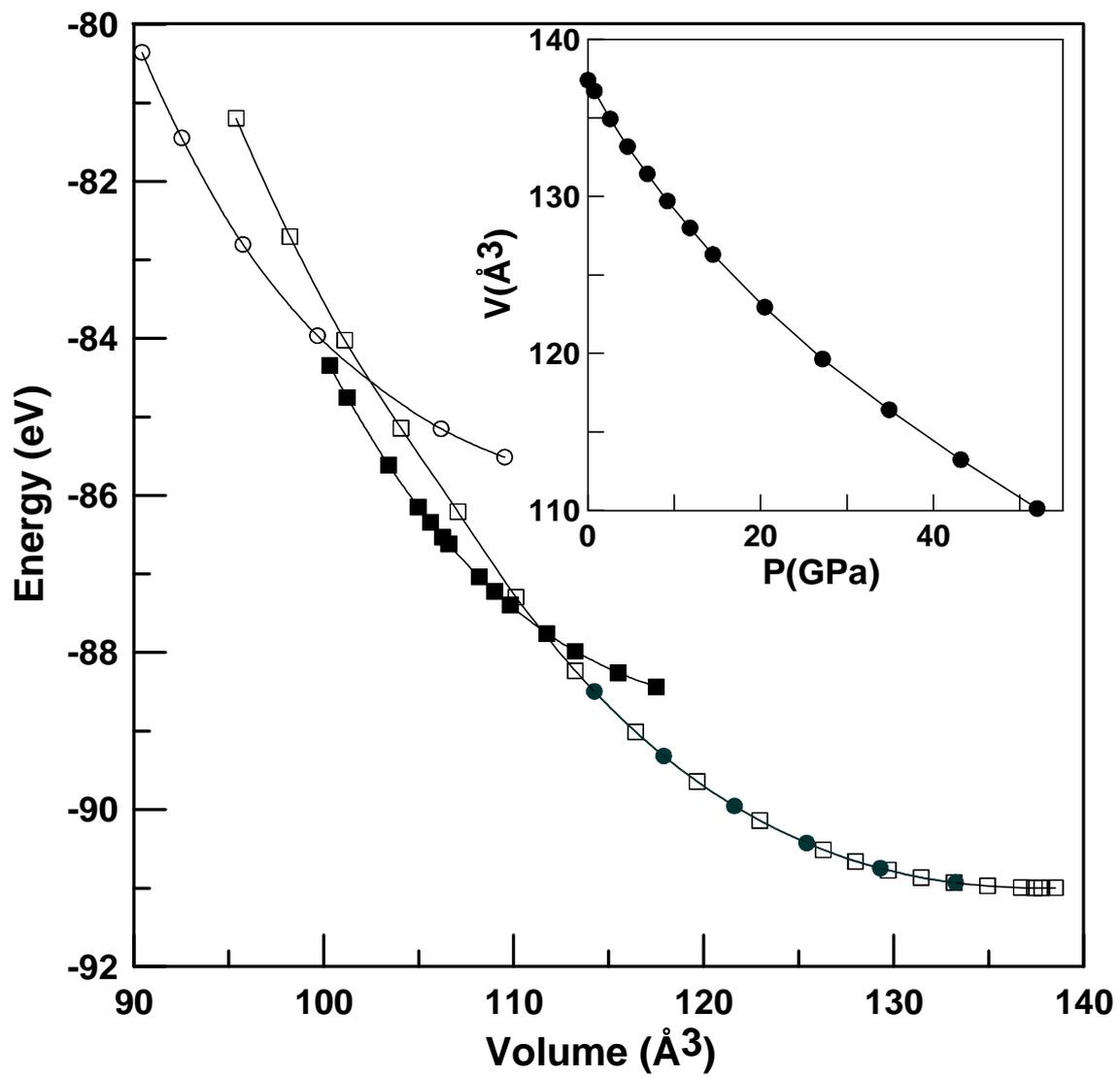

**Figure 5**

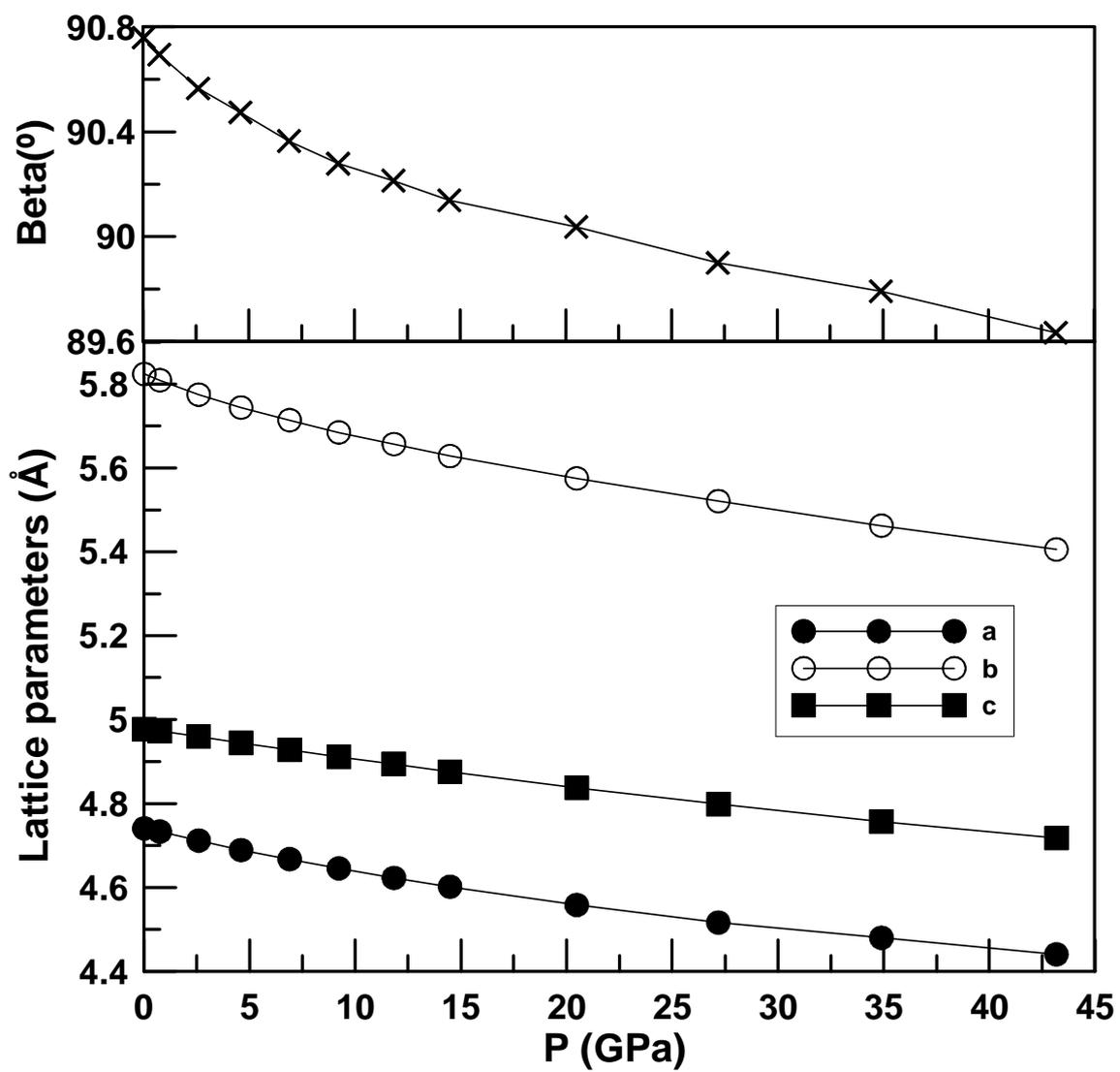

**Figure 6**

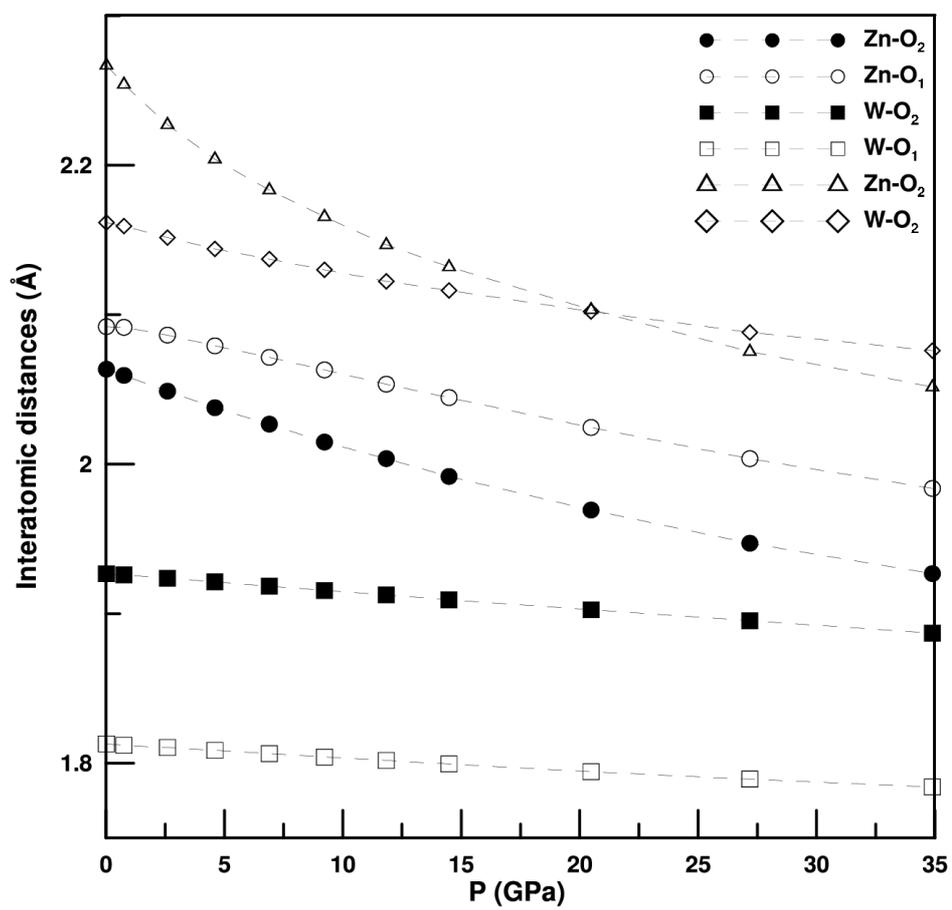